\documentclass[aps,twocolumn,prd,showpacs,amsmath,amssymb]{revtex4}
\usepackage{dcolumn}
\usepackage{bm}
\usepackage{epsf}
\usepackage{amsfonts}
\usepackage{pifont} 

\def\cqg{{\it Class. Quantum Grav.}\ }

\newcommand{\be}{\begin{equation}}
\newcommand{\ee}{\end{equation}} 
\newcommand{\eei}{\end{equation}\indent\indent}
\newcommand{\bc}{\begin{center}}
\newcommand{\ec}{\end{center}}
\newcommand{\ber}{\begin{eqnarray*}}
\newcommand{\ear}{\end{eqnarray*}}
\newcommand{\ba}{\begin{array}}
\newcommand{\ea}{\end{array}}

\newcommand{\bea}{\begin{eqnarray}}
\newcommand{\eea}{\end{eqnarray}}

\newcommand{\ei}{\end{itemize}}

\newcommand{\bra}[1]{\left(#1\right)}
\newcommand{\bras}[1]{\left[#1\right]}
\newcommand{\brac}[1]{\left\{#1\right\}}

\newcommand{\nab}{\nabla}
\newcommand{\la}{\langle}
\newcommand{\ra}{\rangle}

\newcommand \veps {\varepsilon} 

\newcommand{\lb}{\{}
\newcommand{\rb}{\}}
\newcommand{\A}{{\cal A}}
\newcommand{\E}{{\cal E}}
\renewcommand{\H}{{\cal H}}

\newcommand{\lc}{\varepsilon}

\def\case#1/#2{\textstyle\frac{#1}{#2} }

\def\rf#1{(\ref{#1})}

\begin{document}

\title{A new framework for studying spherically symmetric static solutions in  f(R) gravity}

\author{Anne Marie Nzioki$^{ \dag}$, Sante Carloni $^\diamond$, Rituparno Goswami$^{\dag}$
and Peter K.S. Dunsby$^{ \dag \ddag *}$}
\affiliation{* Centre for Astrophysics, Cosmology and Gravitation,  University of Cape Town, Rondebosch,
7701, South Africa}
\affiliation{\dag \ Department of Mathematics and Applied Mathematics, 
University of Cape Town, Rondebosch,
7701, South Africa}
\affiliation{\ddag \  South African Astronomical Observatory, 
Observatory, Cape Town, South Africa}
\affiliation{$\diamond$  Institut d'Estudis Espacials de Catalunya
(IEEC),
Campus UAB, Facultat Ci\`encies, Torre C5-Par-2a pl, E-08193 Bellaterra
(Barcelona) Spain}

\date{\today}

\email{anne.nzioki@gmail.com, sante.carloni@gmail.com, Rituparno.Goswami@uct.ac.za, 
peter.dunsby@uct.ac.za}

\begin{abstract}
We develop a new covariant formalism to treat spherically symmetric spacetimes in 
{\em metric} $f(R)$ theories of gravity. 
Using this formalism we derive the general equations for a 
static and spherically symmetric metric in a general $f(R)$-gravity. These equations are used 
to determine the conditions for which the Schwarzschild metric is the only vacuum 
solution with 
vanishing Ricci scalar. We also show that our general framework provides a clear 
way of showing that 
the Schwarzschild solution is not a unique static spherically symmetric 
solution, providing some incite 
on how the current form of Birkhoff's theorem breaks down for these theories.
\end{abstract}

\maketitle

\section{Introduction}

Ever since the publication of the Schwarzschild solution almost a hundred years ago, the study of spherically symmetric solutions has 
played a fundamental role in determining understanding of the nature of gravity and underlies many of the key tests of Einstein's theory 
of General Relativity. Until recently, General Relativity has been unquestionably the only theory able to explain gravity on both 
astrophysical and cosmological scales. With the advent of new high precision cosmological tests, capable of probing physics at very 
large redshifts, this situation has completely changed and recently the large-scale validity of General Relativity has begun to be 
questioned. This is largely due to the fact that in order to fit the standard model of cosmology, which is based on General Relativity 
coupled to standard matter (baryons and radiation), the introduction of two dark components are needed to achieve a consistent picture. 
Specifically, dark matter is needed to fit the astrophysical dynamics at galactic and cluster scales, while a new ingredient, dubbed dark energy, is required in order to explain the observed accelerated behavior of the Hubble flow. Combining the luminosity distance data of 
Supernovae Type Ia \cite{SN}, the Large Scale Structure \cite{LSS}, the anisotropy of Cosmic Microwave Background \cite{CMB} and Baryon Acoustic Oscillations \cite{BAO} suggest that if we retain General Relativity as the theory of the gravitational interaction, the best fit model is a spatially flat Universe, dominated by cold dark matter (CDM) and Dark Energy (DE) in the form of an effective cosmological constant. Although CDM candidates have not yet been directly detected, there are strong arguments that suggest that CDM has a non-gravitational origin \cite{CDM}. The same is not true for DE. The cosmological constant and coincidence problems together with the fact that there are no convincing DE candidates, seems to suggest that the Concordance model is incomplete, and despite enormous effort over the past few years, this problem remains one of the greatest puzzles in contemporary physics. One of the theoretical proposals that has received a considerable amount of attention recently, is that Dark Energy has a geometrical origin. This idea has been driven by the fact that modifications to General Relativity appear in the low energy limit of many fundamental schemes \cite{stringhe,birrell} and that these modifications lead naturally to cosmologies which admit a Dark Energy like era \cite{SalvSolo,revnostra,Odintsov, Carroll,star2007} without the introduction of any additional cosmological fields. Most of the work on this idea has focused on {\it fourth order gravity}, in which the standard Hilbert-Einstein action is modified with terms that are at most of order four in the metric tensor. 

The features of fourth order gravity have been analyzed with different techniques \cite{Background} and all these studies suggest that these cosmologies can give rise to a phase of accelerated expansion, which is considered to be an important footprint of Dark Energy. In particular, dynamical systems analysis shows that for FLRW models there exist classes of fourth order theories which admit a transient decelerated expansion phase, followed by one with an accelerated expansion rate. The first (Friedmann-like) phase provides a setting during which structure formation can take place, followed by a smooth transition to a DE-like era which drives the cosmological acceleration. 

More recently, the theory of linear perturbations for these models has been developed \cite{fRPerturbations} using the 1+3 covariant approach \cite{Covariant}. A number of important features were found which allows one to differentiate the structure growth scenario from what occurs in General Relativity. Firstly, it was found that the evolution of density perturbations is determined by a {\it fourth order} differential equation rather than a second order one. This implies that the evolution of the density fluctuations contains, in general, four modes rather that two and can give rise to a more complex evolution than the one of General Relativity (GR). Secondly, the perturbations are found to depend on the scale for any equation of state for standard matter (while in General Relativity the evolution of the dust perturbations are scale-invariant). This means that even for dust, the evolution of super-horizon and sub-horizon perturbations are different. Thirdly, it was found that the growth of large scale density fluctuations can occur also in backgrounds in which the expansion rate is increasing in time. This is in striking contrast with what one finds in General Relativity and would lead to a time-varying gravitational potential, putting tight constraints on the Integrated Sachs-Wolfe effect for these models.

Although these results are very encouraging,  there are still some important open problems to be addressed. Of particular interest is the degree to which the physics of fourth order gravity is consistent with {\it both} cosmological and solar system scales, indeed, there has been considerable debate about the short-scale behavior of higher order over the past few years leading to much work on the Newtonian and post-Newtonian limits of these theories \cite{Newtonian}. Consequently, measurements coming from weak field limit tests like the bending of light, the perihelion shift of planets and frame dragging experiments represent critical tests for any theory of gravity. Fundamental to confronting such tests with fourth order gravity is the existence of physically viable spherically symmetric solutions in these theories. The aim of this paper is this therefore twofold. Firstly to determine a set of general results for spherically symmetric spacetimes in $f(R)$ gravity, and secondly to obtain a general procedure for generating solutions of this type. 

The present analysis is based on a powerful extension of the 1+3 covariant approach in which the three spatial degrees of freedom are further decomposed relative to a spatial vector \cite{extension}. In the case of spherical symmetry, this is chosen to be the radial direction. This leads to a larger set of covariant variables with their corresponding equations (evolution, propagation and constraint). Furthermore all the equations are developed in the Jordan frame without resorting to any conformal transformations. 

Unless otherwise specified, natural units ($\hbar=c=k_{B}=8\pi G=1$) will be used throughout this paper, Latin indices run from 0 to 3. The symbol $\nabla$ represents the usual covariant derivative and $\partial$ corresponds to partial differentiation. We use the $-,+,+,+$ signature and the Riemann tensor is defined by
\begin{equation}
R^{a}{}_{bcd}=\Gamma^a{}_{bd,c}-\Gamma^a{}_{bc,d}+ \Gamma^e{}_{bd}\Gamma^a{}_{ce}-\Gamma^e{}_{bc}\Gamma^a{}_{de}\;,
\end{equation}
where the $\Gamma^a{}_{bd}$ are the Christoffel symbols (i.e. symmetric in the lower indices), defined by
\begin{equation}
\Gamma^a_{bd}=\frac{1}{2}g^{ae}
\left(g_{be,d}+g_{ed,b}-g_{bd,e}\right)\;.
\end{equation}
The Ricci tensor is obtained by contracting the {\em first} and the {\em third} indices
\begin{equation}\label{Ricci}
R_{ab}=g^{cd}R_{acbd}\;.
\end{equation}
The symmetrization and the antisymmetrization  over the indexes of a tensor are defined as 
\begin{equation}
T_{(a b)}= \frac{1}{2}\left(T_{a b}+T_{b a}\right)\;,\qquad T_{[a b]}= \frac{1}{2}\left(T_{a b}-T_{b a}\right)\,.
\end{equation}
Finally the Hilbert--Einstein action in the presence of matter is given by
\begin{equation}
{\cal A}=\frac12\int d^4x \sqrt{-g}\left[R+ 2{\cal L}_m \right]\;.
\end{equation}

\section{General equations for fourth order gravity}

In a completely general context, a fourth order theory of gravity is obtained by 
adding terms involving $f(R,R_{ab}R^{ab},R_{abcd}R^{abcd})$ to the standard 
Einstein Hilbert action. However, we know the Gauss-Bonnet term 
$(\mathcal{G}=R^2-4R_{ab}R^{ab}+ R_{abcd}R^{abcd})$ is a total differential in 
four dimensions and hence do not affect the field equations. Hence we can replace 
all linear terms of $R_{abcd}R^{abcd}$ with the other two. Furthermore, if the 
spacetime is highly symmetric, then the variation of the term $R_{ab}R^{ab}$ can 
always be rewritten in terms of the variation of $R^2$ \cite{DeWitt:1965jb,Barth:1983hb}. 
It follows that a {\em sufficiently general} fourth-order Lagrangian for a highly symmetric 
spacetime only contains powers of 
$R$ and we can write the action as 
\be
{\cal A}= \frac12 \int d^4x\sqrt{-g}\left[f(R)+2{\cal L}_m\right]\;,
\label{action}
\ee
where ${\cal L}_m$ represents the matter contribution.

Varying the action with respect to the metric gives the following field equations:
\be
f' G_{ab} = T^{m}_{ab}+ \frac12 (f-Rf') g_{ab}
+ \nabla_{b}\nabla_{a}f'- g_{ab}\nabla_{c}\nabla^{c}f', 
\label{field1}	
\ee
where $f'$ denotes the derivative of the function `$f$' w.r.t the { Ricci scalar and  $T^{m}_{ab}$ is the matter stress energy tensor defined as
\be
T^{m}_{ab} = \mu^{m}u_{a}u_{b} + p^{m}h_{ab}+ q^{m}_{a}u_{b}+ q^{m}_{b}u_{a}+\pi^{m}_{ab}.
\ee 
Here $u^a$ is the direction of a timelike observer, $h_{ab}$ is the projected metric on the 3-space perpendicular to $u^a$. 
Also $\mu^{m}$, $p^m$, $q^{m}$ and $\pi^m_{ab}$ denotes the standard matter density, pressure, heat flux and anisotropic stress respectively}. Equations (\ref{field1}) reduce to the standard Einstein field equations when $f(R) = R$.

\section{1+1+2 Covariant approach}

We know that the 1+3 covariant approach, initially developed by Ehlers and Ellis {\cite{Covariant}} has proven to be a very useful technique in many aspects of relativistic cosmology. The approach has been particularly useful in obtaining a deep understanding of many aspects of relativistic fluid flows, whether it is applied in terms of fully non-linear GR effects or the gauge invariant, covariant perturbation formalism. In cosmology these methods have been applied for example to the formalism and evolution of density perturbations \cite{Perturbations} in the universe and to the physics of cosmic microwave background \cite{CovCMB}. This approach is based on a 1+3 threading decomposition of the spacetime manifold w.r.t a timelike congruence as a splitting of spacetime onto a timelike and a orthogonal three-dimensional spacelike hypersurface. All the essential information in the system is captured in a set of {\em kinematic and dynamic} 1+3 variables that have a well defined physical and geometrical significance. These variables satisfy a set of evolution and constraint equations derived from { the} Bianchi and Ricci identities, forming a closed system of equations for a chosen equation of state describing matter.

A natural extension to the 1+3 approach, optimized for problems which have spherical symmetry, is the 1+1+2 formalism developed recently by Clarkson and Barrett \cite{extension}. In this formalism one first proceeds to the same split of the 1+3 approach and then a further one that isolates a specific spatial direction. This allows us to derive a set of variables that are more advantageous to treat systems with one preferred direction. For example in spherically symmetric system the equation for the 1+1+2 variables are scalar equations and are much simpler than the ones of the 1+3 formalism which are in general tensorial. The 1+1+2 formalism was applied to the study of linear perturbations of a Schwarzschild spacetime \cite{extension} and to the generation of electromagnetic radiation by gravitational waves interacting with a strong magnetic field around a vibrating Schwarzschild black hole \cite{Betschart}.

In the following we give a brief review of these formalisms, before applying it to the specific case of $f(R)$-gravity.

\subsection{Kinematics} 

In (1+3) approach first we define a timelike congruence by a timelike unit vector $u^a$. Then the spacetime is split in the form $R\otimes V$ where $R$ denotes the timeline along $u^a$ and $V$ is the 3-space perpendicular to $u^a$. Then  any vector $X^a$ can be projected on the 3-space by the projection tensor $h^a_b=g^a_b+u^au_b$.

At this point, two derivatives are defined: the vector $ u^{a} $ is used to define the \textit{covariant time derivative} (denoted by a dot) for any tensor $ T^{a..b}{}_{c..d} $ along the observers' worldlines defined by
\be
\dot{T}^{a..b}{}_{c..d}{} = u^{e} \nab_{e} {T}^{a..b}{}_{c..d}~, 
\ee
and the tensor $ h_{ab} $ is used to define the fully orthogonally \textit{projected covariant derivative} $D$ for any tensor $ T^{a..b}{}_{c..d} $ , 
\be
D_{e}T^{a..b}{}_{c..d}{} = h^a{}_f h^p{}_c...h^b{}_g h^q{}_d h^r{}_e \nab_{r} {T}^{f..g}{}_{p..q}~, 
\ee
with total projection on all the free indices. Angle brackets to denote orthogonal projections of vectors and the
orthogonally \textit{projected symmetric trace-free} PSTF part of tensors \cite{Cargese}:
\be
V^{\la a \ra} = h^{a}{}_{b}V^{b}~, ~ T^{\la ab \ra} = \left[ h^{(a}{}_c {} h^{b)}{}_d - \frac{1}{3} h^{ab}h_{cd}\right] T^{cd}\;, 
\ee
In the (1+1+2) approach we further split the 3-space $V$, by introducing the unit vector $ e^{a} $ orthogonal to $ u^{a} $ so that
\be
e_{a} u^{a} = 0\;,\; \quad e_{a} e^{a} = 1.
\ee
Then the \textit{projection tensor}
\be 
N_{a}{}^{b} \equiv h_{a}{}^{b} - e_{a}e^{b} = g_{a}{}^{b} + u_{a}u^{b} 
- e_{a}e^{b}~,~~N^{a}{}_{a} = 2~, 
\label{projT} 
\ee 
projects vectors onto the 2-surfaces orthogonal to $e^{a}$ \textit{and} $u^a$, which, following  \cite{extension}, we will refer to as `{\it sheets}'. Hence it is obvious that $e^aN_{ab} = 0 =u^{a}N_{ab}$. As we know in (1+3) approach any second rank symmetric 4-tensor can be split into a scalar along $u^a$, a 3-vector and a {\it projected symmetric trace free} 
(PSTF) 3-tensor. In (1+1+2) slicing, we can take this split further by splitting the 3-vector and PSTF 3-tensor with respect to $e^a$. Any 3-vector, $\psi^{a}$, can be irreducibly split into a component along 
$e^{a}$ and a sheet component $\Psi^{a}$, orthogonal to $e^{a}$ i.e. 
\be
\psi^{a} = \Psi e^{a} + \Psi^{a}\,, \quad \Psi\equiv\psi^{a} 
e_{a}\,, \quad\Psi^{a} \equiv N^{ab}\psi_{b}~. 
\label{equation1} 
\ee 
A similar decomposition can be done for PSTF 3-tensor, $\psi_{ab}$, which can be split into scalar (along $e^a$), 2-vector and 2-tensor part as follows:
\be 
\psi_{ab} = \psi_{\la ab\ra} = \Psi\bra{e_{a}e_{b} - \frac{1}{2}N_{ab}}+
2 \Psi_{(a}e_{b)} + \Psi_{ab}~, 
\label{equation2} 
\ee 
where 
\bea
\Psi &\equiv & e^{a}e^{b}\psi_{ab} = -N^{ab}\psi_{ab}~,\nonumber \\ 
\Psi_{a} &\equiv & N_{a}{}^be^c\psi_{bc}~,\nonumber \\ 
\Psi_{ab} &\equiv & \psi_{\brac{{ab}}} 
\equiv \bra{ N^{c}{}_{(a}N_{b)}{}^{d} - \frac{1}{2}N_{ab} N^{cd}} \psi_{cd}~,
\eea 
and the curly brackets denote the PSTF part of a tensor with respect to $e^{a}$. \\
We also have 
\be 
h_{\left\{ab\right\}} = 0~,~ N_{\la ab\ra} = -e_{\la a}e_{b\ra} = N_{ab}- \frac{2}{3}h_{ab}~. 
\ee 
The sheet carries a natural 2-volume element, the alternating Levi-Civita 
2-tensor: 
\be
\veps_{ab}\equiv\veps_{abc}e^{c} = \eta_{dabc}e^{c}u^{d}~, \label{perm}
\ee 
where $\veps_{abc}$ is the 3-space permutation symbol the volume element of the 3-space and $\eta_{abcd}$ is  the space-time permutator or the 4-volume. 

With these definitions it follows that any 1+3 quantity  can be locally split in the 1+1+2 setting into only three types of objects: scalars, 2-vectors in the sheet, and PSTF 2-tensors (also defined on the sheet). 
\subsection{Derivatives and the kinematical variables} 

Apart from the `{\it time}' (dot) derivative, of an object (scalar, vector or tensor) which is the derivative 
along the timelike congruence $u^a$, we now introduce two new derivatives, which $ e^{a} $ defines, for any object $ \psi_{a...b}{}^{c...d}  $: 
\bea
\hat{\psi}_{a..b}{}^{c..d} &\equiv & e^{f}D_{f}\psi_{a..b}{}^{c..d}~, 
\\
\delta_f\psi_{a..b}{}^{c..d} &\equiv & N_{a}{}^{f}...N_{b}{}^gN_{h}{}^{c}..
N_{i}{}^{d}N_f{}^jD_j\psi_{f..g}{}^{i..j}\;.
\eea 
The hat-derivative is the derivative along the $e^a$ vector-field in the surfaces orthogonal to $ u^{a} $. The $\delta$ -derivative is the projected derivative onto the sheet, with the projection on every free index.
We can now decompose the covariant derivative of $e^a$ in the direction orthogonal to $u^a$ into it's irreducible parts giving 
\be 
{\rm D}_{a}e_{b} = e_{a}a_{b} + \frac{1}{2}\phi N_{ab} + 
\xi\veps_{ab} + \zeta_{ab}~, 
\ee
where 
\bea 
a_{a} &\equiv & e^{c}{\rm D}_{c}e_{a} = \hat{e}_{a}~, \\ 
\phi &\equiv & \delta_ae^a~, \\  \xi &\equiv & \frac{1}{2} 
\veps^{ab}\delta_{a}e_{b}~, \\ 
\zeta_{ab} &\equiv & \delta_{\lb a}e_{b \rb }~.
\eea
We see that {for an observer } that chooses $ e^{a} $ as special direction in the spacetime, $\phi$ represents the \textit{expansion of the sheet},  $\zeta_{ab}$ is the \textit{shear of $e^{a}$} (i.e. the distortion of the sheet) and $a^{a}$ its \textit{acceleration}. We can also interpret $\xi$ as the \textit{vorticity} associated with $e^{a}$ so that it is a representation of the ``twisting'' or rotation of the sheet. 

Using equations (\ref{equation1}) 
and (\ref{equation2}) one can split the (1+3) kinematical variables and Weyl tensors as
\bea 
\dot{u}^{a} &=& \A e^{a}+ \A^{a}~,\label{1+1+2Acc} \\ 
\omega^{a} &=& \Omega e^{a} +\Omega^{a}~,\\ 
\sigma_{ab} &=& \Sigma\bra{ e_ae_b - \frac{1}{2}N_{ab}} + 
2\Sigma_{(a}e_{b)} + \Sigma_{ab}~, \\ 
E_{ab} &=& \E\bra{ e_{a}e_{b} - \frac{1}{2}N_{ab}} + 
2\E_{(a}e_{b)} + \E_{ab}~, \\
H_{ab} &=& \H\bra{ e_{a}e_{b} - \frac{1}{2}N_{ab}} + 2\H_{(a}e_{b)} + 
\H_{ab}~.\label{MagH} 
\eea 
where $E_{ab}$ and $H_{ab}$ are the electric and magnetic part of the 
Weyl tensor respectively. Therefore the key variables of the 1+1+2 formalism are
\be
\brac{ \Theta, \A, \Omega,\Sigma, \E, \H, 
\A^{a},\Omega^{a}, \Sigma^{a}, \E^{a}, \H^{a}, \Sigma_{ab}, \E_{ab}, \H_{ab} }\,.
\ee 
Similarly, we may split the anisotropic fluid variables $q^{a}$ and $\pi_{ab}$:
\bea
q^{a} &=& Qe^{a} + Q^{a}~, \\
\pi_{ab} &=& \Pi\bras{ e_{a}e_{b} -\frac12 N_{ab}} + 
2\Pi_{(a}e_{b)} + \Pi_{ab}~. \label{Anisotropic}
\eea
The full covariant derivatives of $e^{a}$ and $ u^{a}$ in terms of these variables are given in Appendix \ref{appendix}.

\section{1+1+2 equations for LRS-II Spacetimes} 

Locally Rotationally Symmetric (LRS) spacetimes posses a continuous isotropy group at each point and hence a multi-transitive isometry group acting on the spacetime manifold \cite{EllisLRS}. These spacetimes exhibit locally (at each point) a unique preferred spatial direction, covariantly defined, for example, by either vorticity vector field or a non-vanishing non-gravitational acceleration of the matter fluids. The 1+1+2 formalism is therefore ideally suited for covariant description of these spacetimes, yielding a complete deviation in terms of invariant scalar quantities that have physical or direct geometrical meaning \cite{Gary}. 
{The preferred spatial direction in the LRS spacetimes constitutes a local axis of symmetry and in this case $ e^{a} $ is just a vector pointing along the axis of symmetry and is thus called a 'radial' vector. Since LRS spacetimes are defined to be isotropic, this allows for the vanishing of \textit{all} 1+1+2 vectors and tensors, such that there are no preferred directions in the sheet. Thus, all the non-zero 1+1+2 variables are covariantly defined scalars. The variables, $\brac{\A, \Theta,\phi, \xi, \Sigma,\Omega, \E, \H, \mu, p, \Pi, Q }$ fully describe LRS spacetimes and are what is solved for in the 1+1+2 approach.  A detailed discussion of the covariant approach to LRS perfect fluid space-times can be seen in \cite{EllisLRS}.

A subclass of the LRS spacetimes, called LRS-II, contains all the LRS spacetimes that are rotation free. As consequence in LRS-II spacetimes the variables $\Omega$, $ \xi $ and $ \H $ are identically zero and  the variables  $$\brac{\A, \Theta,\phi, \Sigma,\E, \mu, p, \Pi, Q }$$ fully characterize the kinematics. The propagation and constraint equations for these variables are:

\subsection*{Propagation equations}
\bea 
\hat\phi&= -&\frac12\phi^2+\bra{\frac13\Theta
+\Sigma}\bra{\frac23\Theta-\Sigma}\nonumber\\
&&-\frac23\mu -\frac12\Pi-\E\;, \label{LRS2start}\\
\hat\Sigma-\frac23\hat\Theta&= -&\frac32\phi\Sigma-Q~,
\\
\hat\E-\frac13\hat\mu+\frac12\hat\Pi&= -&\frac32\phi\bra{\E+
\frac12\Pi}\nonumber\\ 
&&+ \bra{\frac12\Sigma-\frac13\Theta}Q~.
\eea

\subsection*{Evolution equations}
\bea
\dot\phi &= -&
\bra{\Sigma-\frac23\Theta}\bra{\A-\frac12\phi} +Q~,\label{Q}
\\
\dot\Sigma-\frac23\dot\Theta&= -&\A\phi +2\bra{\frac13\Theta
-\frac12\Sigma}^{2}\nonumber\\
&&+\frac13 \bra{\mu+3p}-\E +\frac12\Pi~,
\\ 
\dot\E-\frac13\dot\mu+\frac12\dot\Pi&= &\bra{\frac32\Sigma-\Theta}\E
+\frac14\bra{\Sigma-\frac23\Theta}\Pi\nonumber\\&&
+\frac12\phi Q-\frac12\bra{\mu+p}\bra{\Sigma-\frac23\Theta}~. 
\label{LRS2middle}
\eea 

\subsection*{Propagation/Evolution Equations}
\bea
\dot\mu+\hat Q&= -&\Theta\bra{\mu+p}-\bra{\phi+2\A}Q -
  \frac32\Sigma\Pi~,
\\ 
\dot Q +\hat p+\hat\Pi&= -&\bra{\frac32\phi+\A}\Pi-
\bra{\frac43\Theta+\Sigma} Q \nonumber\\
&&-\bra{\mu+p}\A~,
\\
\hat\A-\dot\Theta&= -&\bra{\A+\phi}\A+\frac13\Theta^2 \nonumber\\
&&   +\frac32\Sigma^2 + \frac12\bra{\mu+3p}~. 
\label{LRS2end}
\eea

\subsection*{Commutation relation}
\be
\hat{\dot{\psi}} - \dot{\hat{\psi}} = -\A \dot\psi + 
\bra{\frac13\Theta+\Sigma}\hat{\psi}\;. \label{commutation}
\ee 
Here the quantities $\mu$ and $p$ are the total {\it effective} energy density and 
pressure. In context of fourth order gravity we would define these quantities 
later.
Since the vorticity vanishes, the unit vector field $ u^{a} $ is 
hypersurface-orthogonal to the spacelike 3-surfaces whose intrinsic 
curvature can be calculated from the \textit{Gauss equation} for $ u^{a} $ 
that is generally given as \cite{Gary}:
\be 
^{(3)}R_{abcd}=\bra{R_{abcd}}_{\bot} - K_{ac}K_{bd} + K_{bc}K_{ad}\;,
\ee
where $ ^{(3)}R_{abcd} $ is the \textit{3-curvature tensor}, $ \bot $ 
means projection with $ h_{ab} $ on all indices and $ K_{ab} $ 
is the \textit{ extrinsic curvature}. With the additional constraint of the 
vanishing of the sheet distortion $ \xi $, {\it i.e.} the sheet is a genuine 
2-surface, The Gauss equation for $ e^{a} $ together with the 3-Ricci 
identities determine the 3-Ricci curvature tensor of the spacelike 
3-surfaces orthogonal to $ u^{a} $ to be
\be
^{3}R_{ab} = -\bras{\hat{\phi}+\frac12 \phi^{2}}e_{a}e_{b} - 
\bras{\frac12 \hat{\phi} + \frac12\phi^{2} - K}N_{ab}\;,
\ee
This gives the 3-Ricci-scalar as
\be
^{3}R = -2\bras{\frac12 \hat{\phi} + \frac34\phi^{2} - K} \label{3nRicci}
\ee
where $ K $ is the \textit{Gaussian curvature} of the sheet, $ ^{2}R_{ab}=KN_{ab} $ . 
From this equation and (\ref{LRS2start}) 
an expression for $ K $ is obtained in the form \cite{Gary}
\be
K = \frac13 \mu - \E - \frac12 \Pi + \frac14 \phi^{2} - 
\bra{\frac13 \Theta - \frac12 \Sigma}^{2} \label{GaussCurv}
\ee
From (\ref{LRS2start}-\ref{LRS2middle}), 
the evolution and propagation equations of $K$ can be determined as 
\bea
\dot{K} = -\frac23 \bra{\frac23 \Theta - \Sigma}K,\label{evoGauss}\\ 
\hat{K} = -\phi K. \label{propGauss}
\eea
From equation (\ref{evoGauss}), it follows that whenever the Gaussian curvature 
of the sheet is non-zero and constant in time, then the shear is always 
proportional to the expansion as $ \Sigma=\frac23 \Theta $. 

Let us now turn to the case of spherically symmetric static spacetimes
which belong naturally  to LRS class II. The condition of staticity implies that the dot derivatives of 
all the quantities vanish. Furthermore the expansion also vanishes, as 
a non-vanishing expansion would imply that the timelike congruence would 
contract or expand in time, which is not possible in a static spacetime.
Hence we have $\Theta=0$, and as discussed in the previous section this imply 
$\Sigma=0$. From equation (\ref{Q}) we then have the heat flux $Q$ to vanish 
identically in these spacetimes. Hence the set of (1+1+2) equations which 
describe the spacetime become,
\bea \label{StSpSymEqGen}
\hat\phi &= -&\frac12\phi^2 -\frac23\mu-\frac12\Pi-\E~,
\label{equation1a}\\
\hat\E -\frac13\hat\mu + \frac12\hat\Pi &=-& \frac32\phi\bra{\E+\frac12\Pi}~,
\label{equation2a}\\
0 &= -& \A\phi + \frac13 \bra{\mu+3p} -\E +\frac12\Pi~,
\label{equation3a}\\ 
\hat p+\hat\Pi&= -&\bra{\frac32\phi+\A}\Pi-\bra{\mu+p}\A~,
\label{equation4a}\\
\hat\A &= -&\bra{A+\phi}\A + \frac12\bra{\mu +3p}~. 
\label{equation5a} 
\eea

\section{Spherically symmetric static spacetimes in higher order Gravity}

At this point one can re-derive these equations in the case of $f(R)$-gravity. 
The quantities $\mu$, $p$ and $\Pi$ are defined, in this case, as
\bea
\mu &=& \frac{1}{f'} \left(\mu^m+\frac12 ( Rf' - f) + f''\hat X \right.\nonumber\\
&&\left. + f''X \phi+ f'''X^{2}\right)\\
p &=& \frac{1}{f'} \left(p^m+ \frac12 ( f - Rf') - \frac{2}{3} f''\hat X -\right.\nonumber\\ 
&&\left.\frac{2}{3} f''X \phi - \frac{2}{3} f'''X^{2}- \A f''X\right)\\
\Pi &=& \frac{1}{f'} \bra{\frac23 f''' X^{2} + \frac23 f'' \hat X - \frac13 f'' X \phi}, 
\eea
where we have defined $\hat{R}=X$. We will consider here the ``external" field generated 
by a point-like source so that  $\mu^m=0$ and $p^m=0$.
Because of the additional degrees of freedom the  equations (\ref{equation1a}-\ref{equation5a}) are 
not closed and we have to add an additional equation, the {\it Trace equation}:
\be
Rf'- 2f =  - 3f''\hat X - 3f''X \phi - 3f'''X^{2} - 3\A f''X \;.\label{trace1}
\ee
Using the above equations in (\ref{equation1a}-\ref{equation5a}) and eliminating $\E$,
we get the set of four coupled first order equations governing the spacetime in the fourth 
order gravity as
\bea
&&f'\bras{\hat\phi + \phi\bra{\frac12\phi-\A} }=\frac{1}{3}R f' - \frac23 f \nonumber\\ 
&& +f''X\bra{\phi + 2\A }\;, \label{eq1}\\
&&f'\bras{\hat\A +\A( \A +  \phi)} =  \frac16 f - 
\frac13 R f' -  f''X \A \;,\label{eq2}\\
&&\hat R =  X \;,\\
&&f'' \hat X =  -\frac13 Rf'+ \frac23 f  
- f'''X^{2} - X( \phi + \A )f''  \;.\label{eq4}
\eea
We emphasize here that the above system of equations are written in terms of the 
covariant quantities in the 1+1+2 splitting and absolutely co-ordinate independent. 
Note that the system reduces to the second order system of GR in vacuum \cite{extension}, 
if we put $f(R)=R$, $R=0$ and $X=0$ \footnote{The last two conditions are given by the 
fact that in GR the Ricci scalar is simply proportional to the matter density 
and the pressure of a fluid and becomes automatically zero in vacuum.}. 
However as in the case of the Einstein equation, or any other fully covariant 
system of equations, the physics can be understood fully only if one chooses an 
observer. In the 1+3 approach this is done basically choosing a velocity field,
but in the 1+1+2 framework this is not sufficient. One has to give also a 
particular form of `{\it radial}' co-ordinate. This in turn 
will define a specific form for the `{\it hat}' derivative. As we will see 
in the later  sections  there is a natural choice for this coordinate given 
by the geometry of our problem and we will use it to find 
exact spherically symmetric solutions for some specific $f(R)$ gravity models.

\section{Covariant results for the spherically symmetric system}

From the structure of  (\ref{eq1}-\ref{eq4}) we can {already} deduce
some important results for
spherically symmetric static solutions in {a general} $f(R)$ gravity
in absolute
co-ordinate independent manner. These results are important because they
can be used as guidelines to understand the behavior of any proposed $f(R)$
model in this setting and to design new ones.
\subsection{Necessary condition for existence of solutions with
vanishing Ricci scalar.}
It is evident from the equations (\ref{eq1}-\ref{eq4}) above, the function $f$ must be 
of class $C^3$ at $R=0$,
which imply,
\begin{equation}
|f'(0)|<+\infty\,, \quad |f''(0)|<+\infty\,,|f'''(0)|<+\infty\quad\,.
\end{equation}
Also we impose the conditions
\begin{equation}
f(0)=0, R=0
\end{equation}
Note that the condition of vanishing of the Ricci scalar throughout the manifold
automatically implies $X=0$.

Now there are two possibilities:
\paragraph{ $f'(0)\ne 0$}:
In this case we see the system reduces to the following:
\begin{eqnarray}\label{SchwCond}
&&\hat\phi + \phi\bra{\frac12\phi-\A}=0\;,\\
&&\hat\A +\A( \A +  \phi)=0\;.
\end{eqnarray}
It can be easily checked that the conditions $R=0$, $f(0)=0$ and  $f'(0)\ne 0$ makes 
the Einstein Tensor $G_{ab}$ vanish and therefore Schwarzschild solution is the 
only spherically symmetric static solution.
This then allows us to state a generalization of {\em Birkhoff's Theorem} in
higher order gravity.
\par
{\it For all functions $f(R)$ which are of class $C^3$ at $R=0$ and
$f(0)=0$ while
$f'(0)\ne 0$, Schwarzschild solution is the only static spherically symmetric vacuum 
solution with vanishing Ricci scalar.}
\par
It is also interesting to note  that the above result is consistent with the conditions $f'>0$ and $f''>0$,  
which guarantee  the attractive nature of the gravitational interaction and the absence of tachyons \cite{star2007}.
This shows that there may be a connection between this solution and the very nature
of the gravitational interaction.

The presence of this solution, can have interesting consequences on the
validity of these models on the Solar System level. In particular if one
concludes that the Sun behaves very close to a Schwarzschild solution,
the experimental data of the solar system  would help constraining
these models.
\paragraph{ $f'(0)= 0,f(0)=0$}:
In this case (\ref{eq1}-\ref{eq4}) are identically satisfied for all values
of $\phi$ and $\A$ that guarantees $R=0$ and hence $X=0$ \footnote{It has been noted by several authors that the 
situation $f(0)=f'(0)=0$ is somewhat pathological, since the scalar degree of  freedom of this theory, 
$f'(R)$ corresponds to a Brans-Dicke  scalar field in the equivalent Brans-Dicke  representation, 
with Brans-Dicke parameter $\omega=0$, it also corresponds (apart from a constant) to the inverse 
effective gravitational coupling of the theory. Therefore, $f'=0$  corresponds to infinite gravitational coupling 
$G_{effective}=G/f'$ and  to a singularity of the field equations. However  one can formally set $f' \equiv 0$ and 
look for solutions of the field equations with this constant value of $f'$. A  similar situation has been 
pointed out to occur in scalar-tensor gravity \cite{pathalogical}.}. Hence for all models
with $f'(0)= 0$, any solution with vanishing Ricci Scalar in General Relativity
would be a solution to the above system. This is interesting as it shows
that  fourth order gravity in this context can present the same solutions of
GR plus additional. For example, the
Reissner Nordstrom solution which represent the space time outside a
spherically
symmetric charged body, is a solution to the system (\ref{eq1}-\ref{eq4})
even if no electric charge is present.
Similarly a static spherically symmetric solution for a perfect fluid interior
with equation of state $p=(1/3)\rho$ (for example Hajj-Boutros solution or
the special case of Whittaker solution ~\cite{pf}) can be a solution of this
system in the absence of any standard fluid.
\par
The presence of solutions of type ($b$) shows that when the conditions given
in paragraph ($a$) are not satisfied
the Schwarzschild solution is not a unique static spherically
symmetric solution.
Such results
hint towards a disproof of the general Birkhoff theorem in its classical form 
for fourth order gravity.

\subsection{Necessary condition for existence of solutions with constant scalar
curvature}
Solutions with constant {Ricci scalar} are characterized by the fact that
$R=R_0=const.$ and, as consequence,
$X,\hat{X}=0$. Imposing these conditions on (\ref{eq1}-\ref{eq4})
and supposing it to be at least of class $C^3$ in $R=R_0$ one obtains
\bea
&&f'_0\bras{\hat\phi + \phi\bra{\frac12\phi-\A} }=\frac{1}{3}R_0 f'_0
- \frac23 f_0 \;,\\
&&f'_0\bras{\hat\A +\A( \A +  \phi)} =  \frac16 f_0 -
\frac13 R_0 f'  \;,\\
&&0 =  - R_0f'_0+2 f_0 \;,
\eea
where $f'(R_0)=f'_0$ {\it etc}. A first solution exists if
\be\label{SolConstCurv-a}
f'_0\neq0\;, \quad f_0\neq0\;, \quad 2f_0-R_0f'_0=0\,.
\ee
Instead in the case $f'_0\neq0$, $f_0=0$ one obtains again the
Schwarzschild solution ($R_0=0$).
Finally another solution can be achieved if
\begin{eqnarray}\label{SolConstCurv-b}
&& f'_0=0\;,\quad f_0=0\;, \quad R=R_0\;, \quad X,\: \hat{X} =0\;,
\end{eqnarray}
is satisfied. As in the previous subsection, in this case also, any
constant Ricci scalar solution in GR would identically be a solution
to the system.
The relation \rf{SolConstCurv-a} was already found by Barrow and Ottewill
\cite{bi:Barrow} in the cosmological context and later rediscovered in
\cite{bi:Multamaki}. It relates the value of the constant {Ricci scalar}
with the universal constants in the action. For example if we have the
Lagrangian as $R-2\Lambda$, which is the Lagrangian for GR with the
cosmological constant, we must have, {as well known}, the relation
$R_0=4\Lambda$.
\subsection{The curious case of $R^2$ gravity.}
As we have already explained,
the condition for existence of solutions with {covariantly} constant
scalar curvature {connects} the
constant curvature with the universal constants of the Lagrangian.
However this is
not the case for $f(R)=KR^2$. In fact for this type of Lagrangian the third
condition of (\ref{SolConstCurv-a}) is identically satisfied.
This means that we can have a constant curvature solution for any
value of the curvature.
Thus for $R^2$ gravity, the `{\it cosmological}' constant term in a
Schwarzschild-dS/AdS spacetime becomes
a local constant of integration just like the mass.
{ Hence in this theory we can have two distant stars behaving like two different
Schwarzschild-dS/AdS object with different values of the constant. Unfortunately this
case is rather pathological since it corresponds to the case in which the trace of the field equations in vacuum, 
$3\Box  f' +f'R-2f=0$ is satisfied {\em identically} for constant  Ricci scalar, whereas usually it may be satisfied for special 
values of  $R$. In any case this model is ruled out by solar system experiments (see \cite{R2})}.

\section{Choosing a co-ordinate system and relation between 
the covariant variables and the metric}

The most natural way to choose the proper radial co-ordinate in spherically symmetric static spacetimes, is to make the Gaussian curvature `$K$' of the spherical sheets to be proportional to the inverse {square} of the radius. In that case, this co-ordinate `$r$' becomes the {\it area radius} of the sheets. This gives a geometrical definition to the `{\it hat}' derivative. As we have already seen, $\hat{K} = -\phi K$, therefore the most natural way to define the hat derivative of any scalar $M$ would be
\be
\hat{M}=\frac{1}{2}r\phi\frac{d M}{d r}\,.
\ee
With this choice, the system of equations (\ref{eq1}-\ref{eq4}) becomes
\begin{widetext}
\bea
&&f'\bra{\frac12 r \phi\frac{d\phi}{d r} + \frac12\phi^2-A \phi}=\frac{1}{3}R f' - \frac23 f + 
\bra{\phi + 2\A }f''X\;, \label{eq1a}\\
&&f'\bra{\frac12 r \phi\frac{d\A}{d r} +\A^{2} + \A \phi}  =  \frac16 f - 
\frac13 R f' - \A f'' X  \;,\label{eq2a}\\
&&\frac12 r \phi\frac{d R}{d r}=X\;,\label{eq3a}\\
&&f'' \frac12 r \phi\frac{d X}{d r} =  -\frac13 Rf'+ \frac23 f 
- f'''X^{2} - (X \phi +X \A )f''  \;.\label{eq4a}
\eea
\end{widetext}
In the $r$ co-ordinate above  
the most general spherically symmetric static metric is
\be
ds^{2}= - A(r)dt^{2} + B(r)dr^{2} + r^{2}(d\theta^{2} + \sin^2\theta d\phi^{2})
\ee   
Now, from the properties of the four-velocity $ u^{a} $ and the radial vector 
$ e^{a} $ i.e. $ u^{a}u_{a} = -1 $ and $ e^{a}e_{a}= 1 $, we find that
\be 
u^{t}= \sqrt{A(r)}\;,\;e^{r}= \sqrt{B(r)} \,,
\ee
also, from the definitions of different covariant scalars, we get
\be
\A= - u^{b}u^{a}\nab_{b}e_{a}=\frac{1}{2A}\frac{d A}{d r} \sqrt{B} 
\label{Asolution}
\ee
\be
\phi=N^{b}{}_{a} \nab_{b}e^{a}=\frac2r\sqrt{B} 
\label{Phisolution}
\ee
{$R$ can be found in the usual way as a contraction of the Riemann tensor and $X$ is derived from it as in (\ref{eq3a})}. Thus we see any solution 
to the equations (\ref{eq1a}-\ref{eq4a}), would uniquely determine the metric for 
the spacetime.

\section{An Example: some exact solutions for $R^n$ Gravity}

In this section we present, as an example, a few exact solutions for $R^n$ gravity, in absence 
of standard matter. Specializing the choice of $f(R) = R^{n}$ , equations 
(\ref{eq1a}-\ref{eq4a}) becomes
\bea\label{Rnsys}
\frac 12 n r \phi\frac{d\phi}{d r}R^{n-1} &=&\bra{A - \frac12\phi} \phi R^{n-1} + \frac{n-2}{3n}R^{n}\nonumber\\
&&+ (n-1)R^{n-2}X \bra{\phi +2\A}  \;, \label{eq1b}\\
\frac12 n r \phi\frac{d\A}{d r} R^{n-1}  &=& -\bra{\A + \phi}\A R^{n-1} +  
\frac{1-2n}{6n}R^{n} \nonumber\\
&&-(n-1)R^{n-2}X \mathcal{A}   \;,\label{eq2b}\\
\frac12 r \phi\frac{d R}{d r} &=&  X \;,\\
\frac12 r \phi n(n-1) \frac{d X}{d r} R^{n-2}  &=&  \frac{2-n}{3}R^{n} - X \bra{\phi+\A}
\nonumber\\
&&- n(n-1)(n-2)R^{n-3} X^{2}\,.\label{eq4b}
\eea
\subsection{Schwarzschild solution}
{
Substituting $R=0$,  $dR/ d r=0$ in the above set of equations, we see that 
the equations are satisfied trivially  provided that $n=1,2,\ge3$. 
However since $R=0$ is by itself a 
differential constraint involving $\phi$ and $\A$, hence any $\phi$ and $\A$ 
that ensures a zero Ricci scalar would solve the system. As we know the following 
solution
\be
\phi=\frac{2}{r}\sqrt{1-\frac{2m}{r}}\;,\;\A=\frac{m}{r^2}
\left[1-\frac{2m}{r}\right]^{-\frac12}
\ee
with equations (\ref{Asolution}) and (\ref{Phisolution}) gives the 
usual Schwarzschild metric in $(t,r,\theta,\phi)$ co-ordinates that has a zero 
Ricci scalar, hence the above solution is the solution of the system. 
}
\subsection{A solution with constant non-zero {Ricci scalar}}
As described before if we substitute $X=0$, $R=R_0\ne0$ in the above system of 
equations then a solution is possible if and only if $n=2$. In that case the 
solutions of the other two functions are
\bea
\phi&=&\frac{2}{r}\sqrt{1-\frac{2m}{r}+\frac{R_0}{3}r^2}\\
\A&=&\frac{m+R_0r^2}{r^2}
\left[1-\frac{2m}{r}+\frac{R_0}{3}r^2\right]^{-\frac12}
\eea
This is the usual Schwarzschild-dS/AdS solution depending on the sign of $R_0$.

\subsection{A solution with non-constant {Ricci scalar} vanishing at infinity}

To find more non-trivial solutions of the above system of equations, let us 
use a Schwarzschild like ansatz,
\be
\phi=\sqrt{C_1r^\alpha+C_2r^\beta}\;,\; R=C_3/r^\gamma\; (\gamma>0),
\ee
such that the {Ricci scalar} vanishes at infinity. We use these ansatz 
in the system of equations and then algebraically solve for the powers and 
coefficients such that the system is identically satisfied. 
With this choice we get the following solution:
\begin{widetext}
\bea
\A &=&  \frac{-C \bra{5 - 4 n} r^{ \frac{4 n^{2} - 11 n + 9}{n-2}}+(4 n^{2} - 6 n +2)r^{-1}}{2\bra{2-n}} \bra{\frac{\bra{1 + 2 n - 2 n^{2}} \bra{7 - 10 n + 4 n^{2}} \bra{1 + C r^{- \frac{7 - 10 n + 4 n^{2}}{2-n}}}}{(2 - n)^{2}}}^{-\frac12},\\
\phi &=& \frac{2} {r} \bra{ \frac{\bra{1 + 2 n - 2 n^{2}}\bra{7 - 10 n + 4 n^{2}}}
{\bra{2 - n}^{2} \bra{1 + C r^{- \frac{7 - 10 n + 4 n^{2}}{2-n}}}}}^{-\frac12},  \\
R &=& \frac{6n(n-1)}{\bra{2n \bra{n-1}-1}r^{2}},\\
X &=& - \frac{12n \bra{n-1}}{\bra{2n\bra{n-1}-1}r^{3}}
\bra{\frac{\bra{1 + 2 n - 2 n^{2}}\bra{7 - 10 n + 4 n^{2}}}{\bra{2 - n}^{2} 
\bra{1 + C r^{- \frac{7 - 10 n + 4 n^{2}}{2-n}}}}}^{-\frac12}.
\eea
Now solving for the metric coefficients, we get
\be
A(r)= r^{(2n-2)\frac{(2n-1)}{(2-n)}}+\frac{C}{r^{\frac{(5-4n)}{(2-n)}}}\;\;;\;\; 
\frac{1}{B(r)}= \frac{(2-n)^{2}}{(7-10n+4n^{2})(1+2n-2n^{2}))}\bra{1+\frac{C}
{r^{\frac{(7-10n+4n^{2})}{(2-n)}}}}\;. \label{Bconstant}  
\ee
\end{widetext}
{ This solution was originally found by {Clifton \cite{bi:Clifton}}. The solution reduces 
to Schwarzschild for $n=1$ and valid for $n<{(1+\sqrt{3})/2}$ beyond which the 
metric has unphysical signature. However for $n\in (1,(1+\sqrt{3})/2)$ the Ricci scalar 
is negative and hence the action is only real-valued if $n$ is an even rational number. 
That is in it's lowest form the numerator of the fraction is even. This problem can be 
also be resolved by assuming the absolute value of the Ricci scalar in the action. However 
in that case the Schwarzschild limit at $n=1$ is not possible as $|R|$ is not 
differentiable at $R=0$ and hence doesn't belong to the class $C^3$ functions. 
It is also interesting to note that in spite of the 
Ricci scalar vanishing at infinity this solution is {\it not} asymptotically flat. 
}

\section{Conclusion}
In this paper we have analyzed static spherically symmetric metrics
within the $f(R)$-gravity framework.
Using the 1+1+2 formalism we were able to derive equations describing
these metrics
for a general form of the function $f$ and a pointlike source. These equations have been used to obtain a set
if general conditions for the existence
of certain types of static spherically symmetric solutions.  It is important to note that our system of equations are much simpler than what
one obtains when writing the Einstein Field Equations in terms of metric components and therefore it is much easier to find new solutions
and general covariant results. In particular  the results when $f(0)=0$ and $f'(0)=0$ are much more transparent using this approach.

Our results show that the presence of solutions with constant Ricci
Scalar is influenced by the properties
of the derivatives of the function $f$ up to the third order. This
implies that two $f(R)$ models are indistinguishable in
this framework if differences only arise after the third derivative of
$f$ { at the given value of the Ricci scalar}. Also one can probe in general that a form of
the Birkhoff theorem exists for $f(R)$ gravity only if $f(0)=0$ and
$f'(0)\neq 0$, while in general there is more than
a static and spherically symmetric solution for the field equations.

It is also interesting to note that  the conditions $f'>0$, $f''>0$, which guarantee the attractive nature of the gravitational interaction and the absence of tachyons \cite{star2007} are consistent with the the form of Birkhoff's theorem stated here. 
This suggests a possible link between these conditions and the Birkhoff theorem, something
which definitely deserves further study.

In order to extract observable results the 1+1+2 equations need to be further
specialized choosing a specific form of the radial coordinate. This
is equivalent to the choice of an observer in the 1+1+2 formalism, which,
unlike the 1+3 case, requires not only the specification of a velocity
field but also a specific spatial direction.  The requirement that the
Gauss curvature has an inverse square
dependence offers a natural choice for this coordinate.

Once this is done given any $f(R)$-theory of gravity (and sufficient ingenuity)
one can derive static and spherically symmetric solution(s) for this theory.
We have used a $R^n$-gravity as an example to derive some exact solutions.
As expected from our general considerations, since for this class of
models $f'(0),f(0)=0$,
the system (\ref{eq1}-\ref{eq4}) does not admit
a unique solution and Birkhoff's theorem is violated. 

It is worth to stress, however, that such considerations are limited to the  case of pointlike sources. It is known that the situation can be really different int he case of extended ones \cite{Pechlaner}. Such issues will be treated elsewhere.

As a final comment we would like to point out that if one admits in
the fact that
in this framework the Birkhoff theorem is violated, any Newtonian
limit of a background solution will give, in principle, a different
results \cite{Capozziello:2007ms}. This means that there is no way to
calculate the physical Newtonian potential without knowing the {\it exact } background which
characterizes the entire Universe. This is not surprising because in
these theories the relation between local physics and the rest of the
Universe is much tighter that in General Relativity due to their
relation with Mach's principle.

\appendix 
\section{Useful relations} \label{appendix}
In this appendix we give some useful relation needed for the calculations performed in the text.\\

The full covariant derivatives of $e^{a}$ and $ u^{a}$ are 
\begin{widetext}
\be 
\nab_{a} e_{b} = - \A u_{a}u_{b} - u_{a}\alpha_{b} + 
\bra{\Sigma + \frac13\Theta} e_{a} u_{b} + \bra{ \Sigma_{a} - 
\veps_{ac}\Omega^{c}} u_{b} 
+ e_{a}a_{b} + \frac12\phi N_{ab} + \xi\veps_{ab} + \zeta_{ab}~, 
\label{eFullDerive} 
\ee
\be 
\nab_{a}u_{b} = -u_{a}\bra{\A e_{b} + \A_{b}} + 
e_{a}e_{b}\bra{ \frac13\Theta + \Sigma } + e_{a}\bra{ \Sigma_{b} + 
\veps_{bc}\Omega^{c}}
+ \bra{\Sigma_{a}-\veps_{ac}\Omega^c}e_{b} + N_{ab}\bra{ 
\frac13\Theta - \frac12\Sigma} + \Omega \veps_{ab} + \Sigma_{ab}~. 
\label{uFullDerive} 
\ee 
\end{widetext}
The {covariant time derivative of $ e^{a} $ is given by,} 
\be 
\dot{e}_{a} = \A u_{a} + \alpha_{a}~,~~\mathrm{ where }~~\A = e^{a}\dot{u}_{a}~,
\ee 
and $ \alpha_{a} $ is the component lying in the sheet.
\par
The new variables $a_{a}$, $\phi$, $\xi$, $\zeta_{ab}$ and $\alpha_{a}$ are fundamental objects of the spacetime and their dynamics give us information about the spacetime geometry.
The spatial covariant derivative of a scalar $\Psi$ is defined as
\be 
{\rm D}_{a}\Psi= \hat{\Psi}e_{a} + \delta_{a}\Psi~. 
\ee 
while for any vector $\Psi^{a}$ orthogonal to both $u^{a}$ and $e^{a}$ (i.e. $\Psi^{a}$ lies in the sheet), the various parts of its spatial derivative may be decomposed as follows (Note that a bar on a particular index indicates that the vector or tensor lies in the sheet.):
\bea 
{\rm D}_{a}\Psi_{b} &=& -e_{a}e_{b}\Psi_ca^{c} - 
e_{b}\bras{\frac{1}{2}\phi\Psi_{a} + \bra{\xi\veps_{ac}+ \zeta_{ac}} \Psi^{c} }\nonumber\\ 
&+& e_{a}\hat{\Psi} _{\bar{b}}+ \delta_{a}\Psi_{b}~. 
\eea 
Similarly, for a tensor
$\Psi_{ab}$ (where $\Psi_{ab} = \Psi_{\brac{ab}}$) : 
\bea 
{\rm D}_{a}\Psi_{bc} &=& -2e_{a}e_{(b}\Psi_{c)d}a^{d}+ e_{a}\hat{\Psi}_{bc}+ \delta_a\Psi_{bc}
\nonumber \\ 
&&-2e_{(b}\bras{\frac{1}{2}\phi\Psi_{c)a}+ \Psi_{c)}{}^d\bra{\xi\veps_{ad} 
+ \zeta_{ad}} }. 
\eea

For the Levi Civita 2-tensor we have:
\bea
\veps_{ab}e^{b} &=& 0 = \veps_{(ab)}~, \\ 
\veps_{abc} &=& e_{a}\veps_{bc} -e_{b}\veps_{ac} + e_{c}\veps_{ab}~, \\
\veps_{ab}\veps^{cd} &=&  N_{a}{}^{c}N_{b}{}^{d} - N_{a}{}^dN_{b}{}^{c}~, \\ 
\veps_{a}{}^{c}\veps_{bc} &=& N_{ab}~,~~ \veps^{ab}\veps_{ab} = 2~. 
\eea 
and
\bea 
\dot{\lc}_{ab} &=& -2u_{[a}\veps_{b]c}\A^c + 2e_{[a}\epsilon_{b]c}
\alpha^c~, \nonumber \\
\hat{\lc}_{ab} &=& 2e_{[a}\lc_{b]c}a^c~,\nonumber \\
\delta_c\veps_{ab} &=& 0~. 
\eea

For the projection tensor we have
\bea
\dot{N}_{ab}& =& 2u_{(a}\dot{u}_{b)} - 2e_{(a}\dot{e}_{b)} = 
2u_{(a}\A_{b)} - 2e_{(a}\alpha_{b)}~, \nonumber \\
\hat{N}_{ab} &=& -2e_{(a}a_{b)}~, \nonumber \\ 
\delta_{c}N_{ab} &=& 0~. 
\eea 

\section*{Acknowledgments}
SC was funded by Generalitat de Catalunya through the Beatriu de Pinos contract 2007BP-B1 00136. We thank the National Research Foundation (South Africa) for financial support and Timothy Clifton for useful comments. The University of Cape Town provided support by grant for RG and the National Astrophysics and Space Science Program supported AMN.

\end{document}